\documentclass{osa-article}

\journal{josab}

\articletype{Research Article}

\usepackage{placeins}
\usepackage{diagbox}
\usepackage{array}

\newcolumntype{P}[1]{>{\centering\arraybackslash}p{#1}}

\begin{document}

\title{Excitation of E1-forbidden Atomic Transitions with Electric, Magnetic or Mixed Multipolarity in Light Fields Carrying Orbital and Spin Angular Momentum}

\author{Maria Solyanik-Gorgone,\authormark{1*} Andrei Afanasev,\authormark{1} Carl E. Carlson, \authormark{2} Christian T. Schmiegelow, \authormark{3,4} and Ferdinand Schmidt-Kaler \authormark{3}}

\address{\authormark{1}Department of Physics, The George Washington University, Washington, DC 20052, USA\\
\authormark{2}Department of Physics, The College of William and Mary in Virginia, Williamsburg, VA 23187, USA\\
\authormark{3}QUANTUM, Institut f\"ur Physik, Universit\"at Mainz, Staudingerweg 7, 55128 Mainz, Germany\\
\authormark{4}Departamento de Fisica, FCEyN, UBA and IFIBA, Conicet, Pabellon 1, Ciudad Universitaria, 1428 Buenos Aires, Argentina}

\email{\authormark{*}msolyanik@gwmail.gwu.edu} 



\begin{abstract}
Photons carrying a well-defined orbital angular momentum have been proven to modify spectroscopic selection rules in atomic matter.
Excitation profiles of electric quadrupole transitions have been measured with single trapped $^{40}$Ca$^+$ ions for varying polarizations.
We further develop the photo-absorption formalism to study the case of arbitrary alignment of the beam's optical axis with respect to the ion's quantization axis and mixed multipolarity. 
Thus, predictions for $M1$-dominated $^{40}\text{Ar}^{13+}$, E3-driven $^{171}\text{Yb}^+$ and $^{172}\text{Yb}^+$, and B-like $^{20}\text{Ne}^{5+}$ are presented. The latter case displays novel effects, coming from the presence of a strong photon -- magnetic dipole coupling.  
\end{abstract}


\section{Introduction}

The interest of scientists to vortex-like dislocations in various media goes back to 1970's \cite{nye1974dislocations, wesfreid1984cellular}. For historic and topical reviews we would like to refer the reader to such sources as \cite{1464-4258-11-9-090201, yao2011orbital}. The ability of twisted photon beams to transfer Orbital Angular Momentum (OAM) as an extra degree of freedom, first shown in \cite{Allen1992zz}, has largely influenced our understanding of light, e.g. \cite{van1994spin, sztul2006double, bliokh2015spin, milione2011higher, Surzhikov15, Schmiegelow2016, quinteiro2017twisted, padgett2017orbital}. It made OAM-carrying laser beams a popular research subject in optical and quantum communications and information security~ \cite{brunet2016optical, krenn2016twisted, mair2001entanglement, fickler2012quantum, krenn2017orbital}. Non-trivial topological structure of such OAM-carrying optical vortices, paired with progress in photonics and nanotechnology~\cite{monticone2017metamaterial}, resulted in proposals and subsequent implementations of novel devices and techniques~\cite{swartzlander2001peering, maurer2011spatial, trichili2016optical, Yao11, barnett2016optical, Padgett2015}.

Over years, significant progress has been made towards understanding the mechanisms of twisted light interaction with matter. It became clear that in order to detect transfer of photon OAM to the internal degrees of freedom of electron configurations, one needs to consider transitions driven by contributions higher than the electric dipole $E1$. This is the non-trivial electromagnetic field distribution in the beam vortex, that bound electrons with high angular momentum are sensitive to \cite{veysi2016focused, quinteiro2017formulation}. It has been shown for molecular \cite{babiker2002orbital} and atomic \cite{scholz2014absorption} systems, as well as semiconductor heterostructures \cite{quinteiro2010electronic, quinteiro2017magnetic}.

In this paper we focus our attention on OAM-photons interacting with single trapped ions. The possibility to enhance and control weak atomic transitions was discussed in \cite{afanasev2018E2M1}, where the authors foreseen OAM-modes being used is high precision atomic spectroscopy. Particular interest in the field has been payed to light-matter interactions in the case of beams carrying a well-defined OAM~\cite{van1994selection, franke2017optical, peshkov2017photoexcitation, afanasev2018experimental}. In the past a detailed experimental study demonstrated the transfer of photon OAM to the valence electron of a single trapped $^{40}$Ca$^+$ ion~\cite{Schmiegelow2016, afanasev2016high}. This progress was based on the technologies developed for ion-trap quantum computing~\cite{haffner2008quantum}. 

Now we propose extending these methods to include atomic transitions in ions with complex level structure such as Yb$^+$. We also study selection rules in highly charged ions (HCI) as Ar$^{13+}$ and Ne$^{4+}$. We aim to stimulate experimental studies on these transitions with OAM light, and have chosen cases that have already been experimentally studied using Gaussian beams~\cite{schmoger2015coulomb, lu2000commercial, chen2004experimental} or have accessible optical transitions. The theory considers arbitrary alignment of the field vector with respect to the atomic quantization axis and takes hyperfine splitting into account, extending previous investigations~\cite{afanasev2018experimental}.

The paper is structured as follows: the theoretical approach is described in section II. It is built on the  formalism for Bessel (BB) and Bessel-Gauss (BG) presented in \cite{afanasev2016high, afanasev2018experimental, afanasev2018E2M1}, where multipole expansion of the plane-wave contribution has been used, similar to \cite{Surzhikov15}. In section III we compare our findings with data obtained in $^{40}$Ca$^+$ ions as a test case for our theory. In this case next-to-the-leading $M3$ multipole is strongly suppressed, compared to the dominating $E2$. Previously, the authors in \cite{Schmiegelow2016} used semi-classical formalism, while we follow purely quantum approach of evaluating the interaction strength in terms of photo-absorption transition amplitudes. While certain data we use here were already published before, the data for various photon polarization states are presented to the public for the first time. Section IV is devoted to theoretical predictions for the ions with strong electric, magnetic and mixed multipolarities. We study the $M1$-dominant  $P_{1/2} \rightarrow P_{3/2}$ transition in  $^{40}\text{Ar}^{13+}$, the E3-driven transition in $^{171}\text{Yb}$ or $^{172}\text{Yb}$ \cite{olmschenk2007manipulation, welinski2016high, ortu2018simultaneous, wcis?o2018First}  and the mixed $M1+E2$ multipolarity transition in $^{20}\text{Ne}^{5+}$  \cite{rynkun2012energies}. The summary is provided in section V.

\section{Formalism} \label{Sec.2}

For twisted photons in a Bessel beam whose vortex line is displaced from the center of the target atom by a distance $b$, and where the atomic states are quantized along an axis parallel to the beam propagation direction, the transition amplitude $M$ reads \cite{scholz2014absorption, afanasev2018experimental}
\begin{equation}
\begin{aligned}
M_{m_f m_i \Lambda}^{(\text{BB})} (b; \theta_z=0) =& A \;i^{m_f-m_i-2m_{\gamma}} e^{i(m_{\gamma}+m_i - m_f) \phi_b} J_{m_{\gamma}-m_f+m_i}(\kappa b) \times && \\ & \times \sum_{m'_f m'_i} d_{m_f, m'_f}^{j_f}(\theta_k) d_{m_i, m'_i}^{j_i}(\theta_k) M_{m'_fm'_i  \Lambda}^{\text{(pw)}}(0)	&&
\label{29/09/17_2}
\end{aligned}
\end{equation}
This amplitude is proportional to experimentally measured transition strength $\Omega_R = \Omega_{Rabi}$, e.g. \cite{Schmiegelow2016}. $A$ denotes an overall normalization factor. The initial and final atomic states have quantum numbers $\{ j_i , m_i \}$ and $\{ j_f , m_f \}$   (where $j_{i,f}$ are total angular momenta of the electronic states, with orbital angular momenta $l_{i,f}$), and $m_\gamma$ is the projection of the photon beam's total angular momentum on the direction of the beam propagation, the $z$-axis. $J_m(\kappa b)$ is a Bessel function with the argument given by the impact parameter $\pmb{b}=\{b \cos \phi_b, b \sin \phi_b, 0\}$, and the transverse momentum magnitude $\kappa = \sqrt{k^2-k_z^2}$. The $d_{m, m'}^{j}(\theta_k)$ are Wigner rotation matrices evaluated at the pitch angle $\theta_k$, \cite{varshalovich1988quantum}.  $M^{(\text{pw})}$ is the transition amplitude for a plane wave traveling in the $z$-direction with helicity $\Lambda= \pm 1$. We will also define $\ell_{\gamma}$ from the identity $m_{\gamma} \equiv \ell_{\gamma} +\Lambda$, that can be interpreted as photon's OAM in paraxial approximation.

In our practical considerations, we have used a Bessel-Gauss mode, which is a Bessel beam with peripheral behavior suppressed by a Gaussian factor to better fit experimental conditions.  Additionally, when the static magnetic field that defines the quantization axis of the atomic states points in a general direction $\{0, \theta_z, \phi_z\}$, one has
\begin{equation}
\begin{aligned}
M_{m_f m_i \Lambda}^{(\text{BG})} (b; \theta_z) =& e^{-b^2/\text{w}_0^2}e^{-i(m_f-m_i)\phi_z} \times && \\ & \times \sum_{m'_f m'_i} d_{m_f m'_f}^{j_f}(\theta_z) d_{m_i m'_i}^{j_i}(\theta_z) M_{m'_f, m'_i \Lambda}^{(\text{BB})} (b; \theta_z=0)
\label{01/24/2017/2}
\end{aligned}
\end{equation}

\begin{figure}[htbp]
\centering
\includegraphics[scale=0.23]{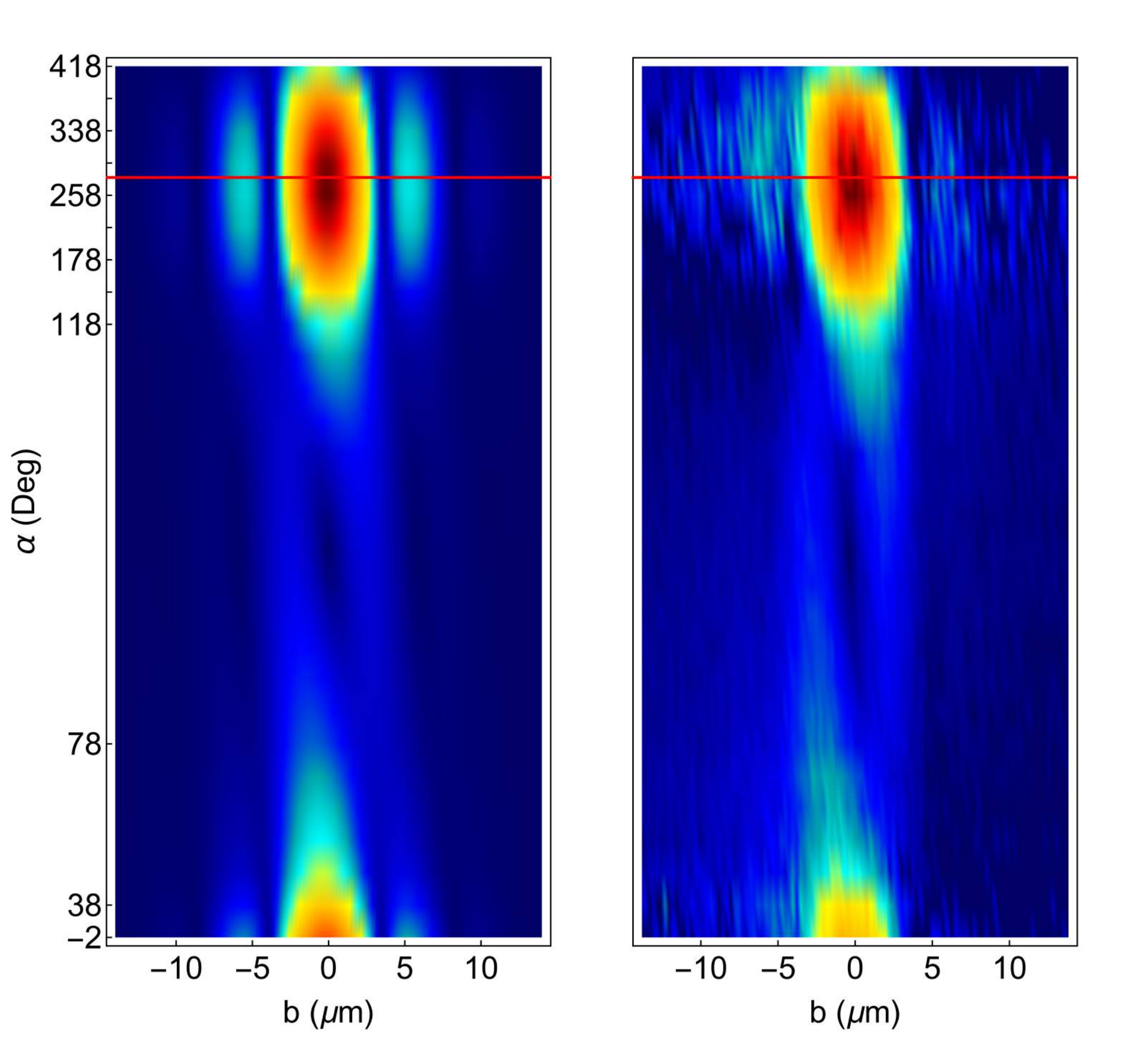}\hspace{8mm}
\includegraphics[scale=0.242]{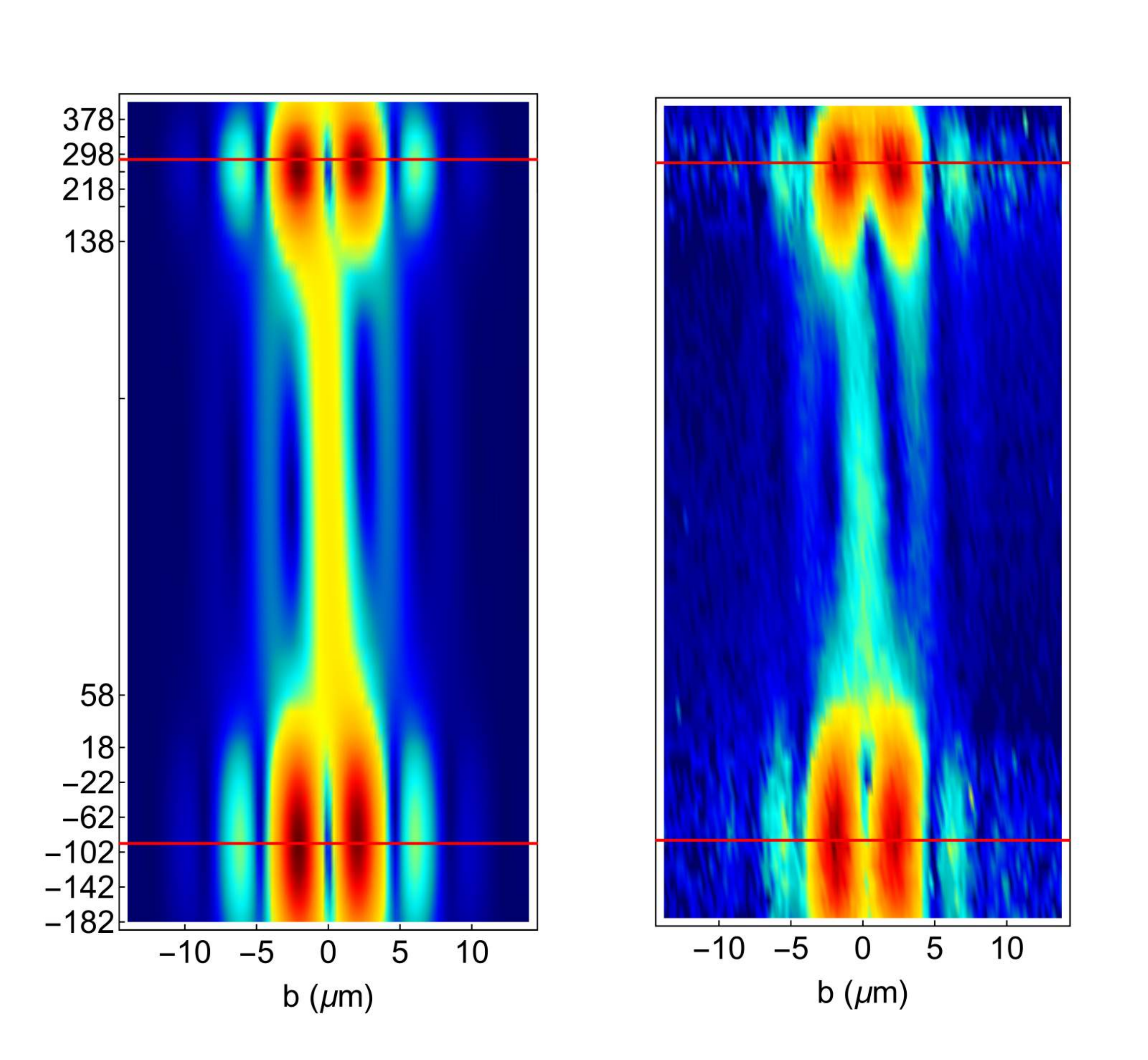}\\
\vspace{3mm}
\includegraphics[scale=0.3]{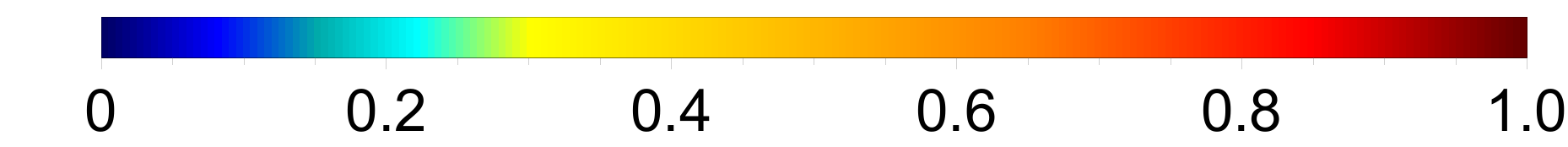}
\caption{(Colors online) Contour-plots of the normalized transition strength as a function of polarization and impact parameter for $4^2S_{1/2} \rightarrow 3^2D_{5/2}$ in a single $^{40}$Ca$^+$. From left to right, first and second subplots are for $\ell_{\gamma} = 0$ theory and experimental data correspondingly. Third and fourth are for $\ell_{\gamma} = 1$ theory and experimental data correspondingly. The theory prediction is fitted using $\theta_k =[ 0.075,0.095]$~rad and $\phi_b = [-0.62,-0.3]$~rad for $\ell_{\gamma} = 0 \text{ and }1\;$respectively. Red lines indicate pure vertical polarization. 
}
\label{11/21/2017/1}
\end{figure}
\hspace{-3.5mm}where w$_0$ is the gaussian width of the beam (beam waist), as defined in e.g. \cite{siegman1986university}.

A plane wave photon can be expanded in terms of spherical waves.  Each spherical wave has total angular momentum $j$, and the plane-wave amplitude can be expressed in terms of spherical multipoles~\cite{afanasev2018E2M1},
\begin{equation}
M_{m_fm_i \Lambda}^{\text{(pw)}}(0) = - \sum_{j=1}^{\infty} i^{j+\mu} \sqrt{\frac{ 4\pi (2j+1) }{ 2j_f+1 }} \Lambda^{\mu+1} C_{j_i\; m_i\; j\; \Lambda}^{j_f\; m_f}  \;  M_{j\mu}
\label{10/5/17_1}
\end{equation}
The $C_{j_i\; m_i; j\; \Lambda}^{j_f\; m_f}$ are Clebsch-Gordan coefficients, and $M_{j\mu}$ is the spherical amplitude of order $j$ and multipolarity $\mu$.  Magnetic multipoles are described with   $\mu = 0$ and electrical ones with $\mu = 1$. We conveniently simplify by writing $M_{j0} = \text{M\,j}$ and $M_{j1} = \text{E\,j}$.  

While the sum in Eq.~\eqref{10/5/17_1} is formally infinite, typically only two amplitudes contribute for given atomic states, and often one of them is small enough to neglect. When an atom has nuclear spin $\pmb{I}$ magnetic hyperfine interaction may significantly contribute to the atom photo-excitation. In case of non-deformed nucleus, such as $^{171}\text{Yb}^{+}$, nuclear electric quadrupole does not contribute and the photon angular momentum couples to the total atomic angular momentum $\pmb{F} = \pmb{J}+\pmb{I}$ (TAM). Hence, in eqn.~\eqref{10/5/17_1} one requires the substitution of the principal quantum numbers $| j_f m_f \rangle \rightarrow | j_f I\; F_f m^{(f)}_f \rangle$ and $| j_i m_i \rangle \rightarrow | j_i I\; F_i m^{(f)}_i \rangle$, e.g. \cite{rodrigues2016excitation}, and of the expansion coefficients
\begin{equation}
\begin{aligned}
\sqrt{\frac{1}{2j_f+1}}C_{j_i\; m_i\; j\; \Lambda}^{j_f\; m_f} \to
(-1)^{j_f + I + F_i - j} \sqrt{(2F_i+1)} \;C_{F_i\; m^{(f)}_i\; j\; \Lambda}^{F_f\; m^{(f)}_f} 
	\left \{ \begin{matrix}
	j_f && F_f && I \\
	F_i && j_i && j \\
	\end{matrix}\right \}
\label{06/06/2018_1}
\end{aligned}
\end{equation}
where $6j$-Wigner coefficients, as defined in \cite{edmonds1957angular}.

\section{$^{40}$Ca$^+$: theory and experiment} \label{sec.3}

We consider the atomic transition $4^2S_{1/2} \rightarrow 3^2D_{5/2}$ in a single trapped $^{40}$Ca$^+$ ion. The coupling scheme, outlined in \eqref{10/5/17_1} and \eqref{06/06/2018_1}, allows two multipole contributions: $M3$ and $E2$.   However, $M3$ 

\begin{figure*}[!ht]
\centering
\vspace{-0.cm}
\centering
\includegraphics[scale=0.29]{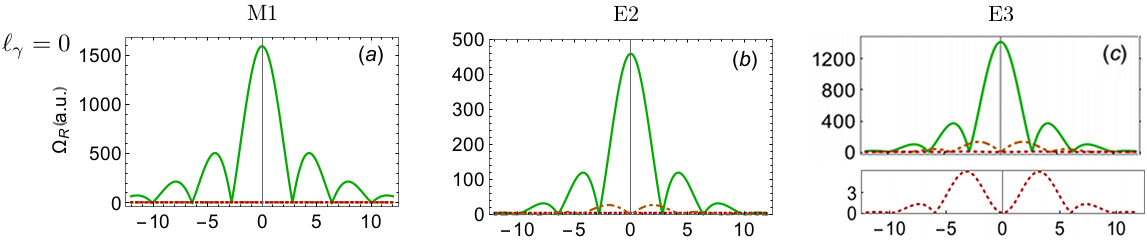}\\ \vspace{1mm}
\includegraphics[scale=0.29]{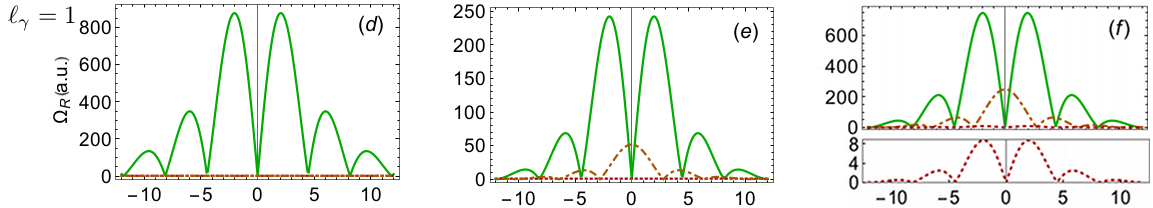}\\ \vspace{1mm}
\includegraphics[scale=0.29]{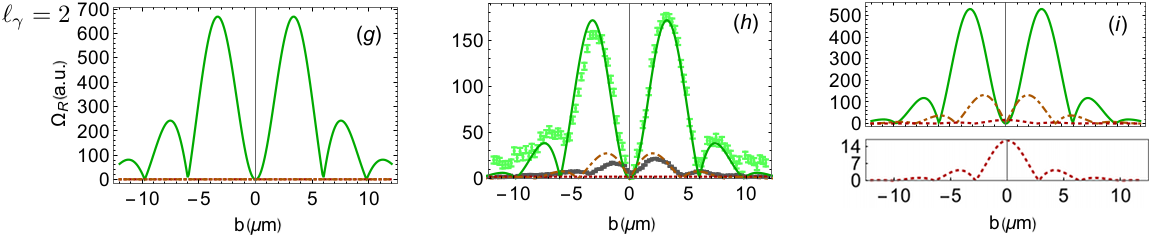}
\vspace{-0.cm}
\centering
\caption{(Colors online) Transition strength $\Omega_r$ for different atomic multipolarities and beam types as a function of impact parameter $b$. The beam direction chosen to be collinear with the external static magnetic field: $\theta_z=0$. Under these conditions conservation of the projection of angular momentum along $z$ is conserved ($\Delta m=\Lambda+\ell_\gamma$) for $b=0$. We choose for all plots left-circular polarized photons $\Lambda=+1$. Columns group multipolarities ($M1, \;E2, \;E3$), while rows group results of increasing orbital angular momentum $\ell_{\gamma} = 0,1,2$. Line-type indicates the magnetic transition $\Delta m = 1 ,2, 3$ for solid-green, dash-dot-brown and dot-red. For magnetic dipole transitions only $\Delta m=1$ transitions are present and there is a maximum at the center only for $\ell_\gamma=0$ where the conservation of projection of angular momentum is conserved. For electric quadrupole ($E2$) transitions, also $\Delta m=2$ is allowed when $\ell_\gamma=1$, as seen in subplot (e). There is a non-zero interaction at the dark center of the beam. Similarly, for the electric octupole ($E3$), $\Delta m=3$ transitions are allowed. As in all cases, transitions are allowed for b=0 only when the conservation rule is met: $\Delta m=\Lambda+\ell_\gamma$. All plots for $E3$ transitions also show and expanded inset with the detail if behavior for $\Delta m=3$ . In figure (h) we contrast our predictions to the results presented in~\cite{Schmiegelow2016}.}
\label{fig:plotsRandL1}
\vspace{-0.cm}
\end{figure*}

\hspace{-3.5mm}is much weaker that $E2$. This is due to $M3$ contributing in the second-to-the-leading order in the Taylor expansion of $e^{i \pmb{k} \cdot \pmb{r}}$, while $E2$ arises in first, along with $M1$, \emph{e.g.} \cite{condon1951theory}. To the best of the authors' knowledge, there are also no spectroscopic effects related to the presence of $M3$. So the plane-wave amplitude $M^{\text{(pw)}}$ can be written as: 
\begin{gather}
M^{\text{(pw)}}_{m_f\;m_i\;\Lambda} \simeq C_{1/2\;m_i\;2\;\Lambda}^{5/2\;m_f} E2
\end{gather}
The interaction strength for a beam of definite helicity $\Lambda$ and TAM $m_{\gamma}$ can then be calculated using equation~\eqref{01/24/2017/2}. One obtains 
\begin{equation}
\begin{aligned}
M_{m_f m_i \Lambda}^{(\text{BG})} (b; \theta_z=0) \simeq & \;A \;i^{m_f-m_i-2m_{\gamma}} e^{-b^2/\text{w}_0^2}  e^{i(m_{\gamma}+m_i - m_f) \phi_b}J_{m_{\gamma}-m_f+m_i}(\kappa b)  \times && \\
&\times \sum_{m'_i} d_{m_f, m'_i+\Lambda}^{j_f}(\theta_k) d_{m_i, m'_i}^{j_i}(\theta_k) C_{j_i\;m'_i\;2\;\Lambda}^{j_f\;m'_i+\Lambda} E2&&
\label{01/10/18_1}
\end{aligned}
\end{equation}

We now compare this results with interaction strengths measured with a single trapped $^{40}$Ca$^+$ ion, Zeeman-spit in static magnetic field \cite{Schmiegelow2016}. In the experiment an ion was trapped in microstructured segmented Paul trap with the 5nm position resolution. The interaction strength 


\begin{figure*}[ht]
\begin{center}
  \includegraphics[scale=0.29]{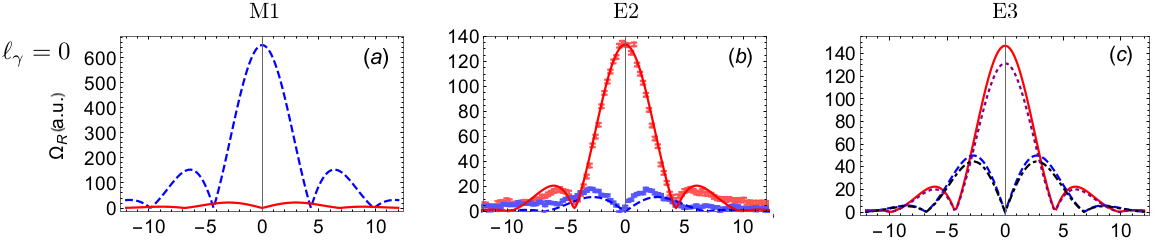}\\ \vspace{1mm}
\includegraphics[scale=0.29]{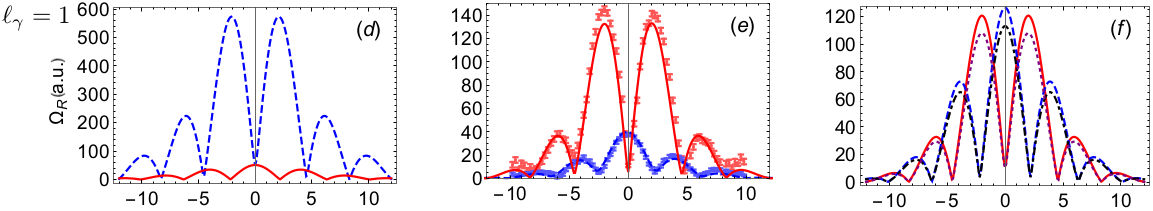}\\ \vspace{1mm}
\includegraphics[scale=0.29]{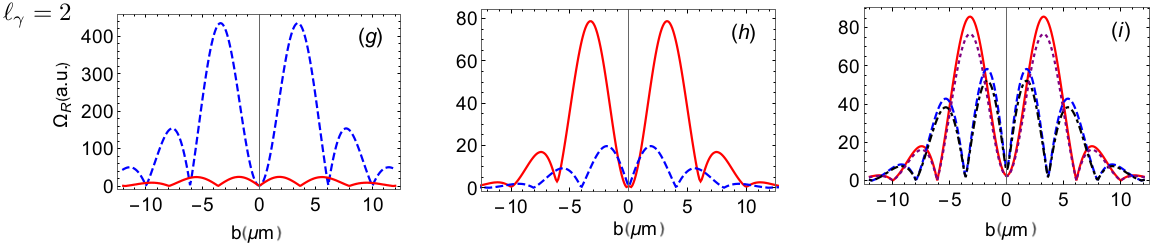}
\end{center}
  \caption{(Colors online) Transition strength $\Omega_r$ for different atomic multipolarities and beam types as a function of impact parameter $b$. 
In all cases the atomic change in magnetic number is $\Delta m=1$. Columns group multipolarities ($M1,\;E2,\;E3$), while rows group results of increasing orbital angular momentum $\ell_{\gamma} = 0,1,2$. 
Red-solid (blue-dashed) indicates results for light polarized vertically (horizontally). For $E3$ we also show the corresponding results for non-null nuclear spin ($\pmb{I} = 1/2$) in dotted-purple (dash-dot-black). We observe the inclusion of nuclear spin in the $E3$ transition produces only a difference in overall strength but no change in structure. Electric transitions ($E2$ and $E3$) are for alignment angle $\theta_z=\pi/4$; they show qualitatively similar behavior.  In subplots (b) and (e) we contrast our predictions with experimental data from~\cite{Schmiegelow2016}.  The magnetic $M1$ transition is shown at alignment angle $\theta_z=\pi/2$ where it also shows qualitatively similar behavior to the electric quadrupole and octupole but with the polarizations inverted.}
\label{fig:plotsHandV}
\end{figure*}

\hspace{-3.5mm}was measured as a Rabi frequency as a function of the position of the ion with respect to the beam (the impact paramer $b$). For the details about the experimental setup and trapping techniques we refer the interested reader to e.g. \cite{Schmiegelow2016}.

First we consider data taken with angle between the quantization axis and beam propagation vector $\theta_z = \pi/4$. We consider two datasets. One, shown in figure~\ref{11/21/2017/1}, where excitation profiles are measured as a function of the a varying polarization. And a second one, shown in figure~\ref{fig:plotsHandV} (d, e), where detailed profiles are measured for two incident linear polarizations: horizontal (H) and vertical (V) to the incidence plane formed by the direction of static magnetic field and the beam axis.

For both datasets on figures~\ref{11/21/2017/1} and~\ref{fig:plotsHandV} we consider the transitions $^2S_{1/2} \rightarrow \; ^2D_{5/2}$ between the Zeeman-split levels: $m_i=1/2$ into $m_f = 3/2$. The polarization angles for the measurements were extracted by comparing the optical response in experiments and those generated in the simulation. We describe the polarization vector of the beam as:
\begin{equation}
\hat{e} = e^{i \delta} (\hat{e}_{-} \cos (\alpha/2)- \hat{e}_{+} \sin (\alpha/2) \;e^{-i 2 \delta})
\label{04/14/2018_1}
\end{equation}
where the polar angle $\alpha$ and phase retardation $\delta$ ranges in the interval $[0, \pi]$. The convention used here for right-circular $\hat{e}_{-}$ and left-circular $\hat{e}_{+}$ polarization states is $\hat{e}_{\pm} = \tfrac{1}{\sqrt{2}} \{ \mp1, -i, 0 \}$.   The H state is obtained with $\alpha = 90^\circ$ and $\delta=0$, and the V state is obtained using $\alpha = 90^\circ$ and $\delta=90^\circ$.

\begin{figure}[h!]
\center
\vspace{-1mm}
\includegraphics[scale=0.44]{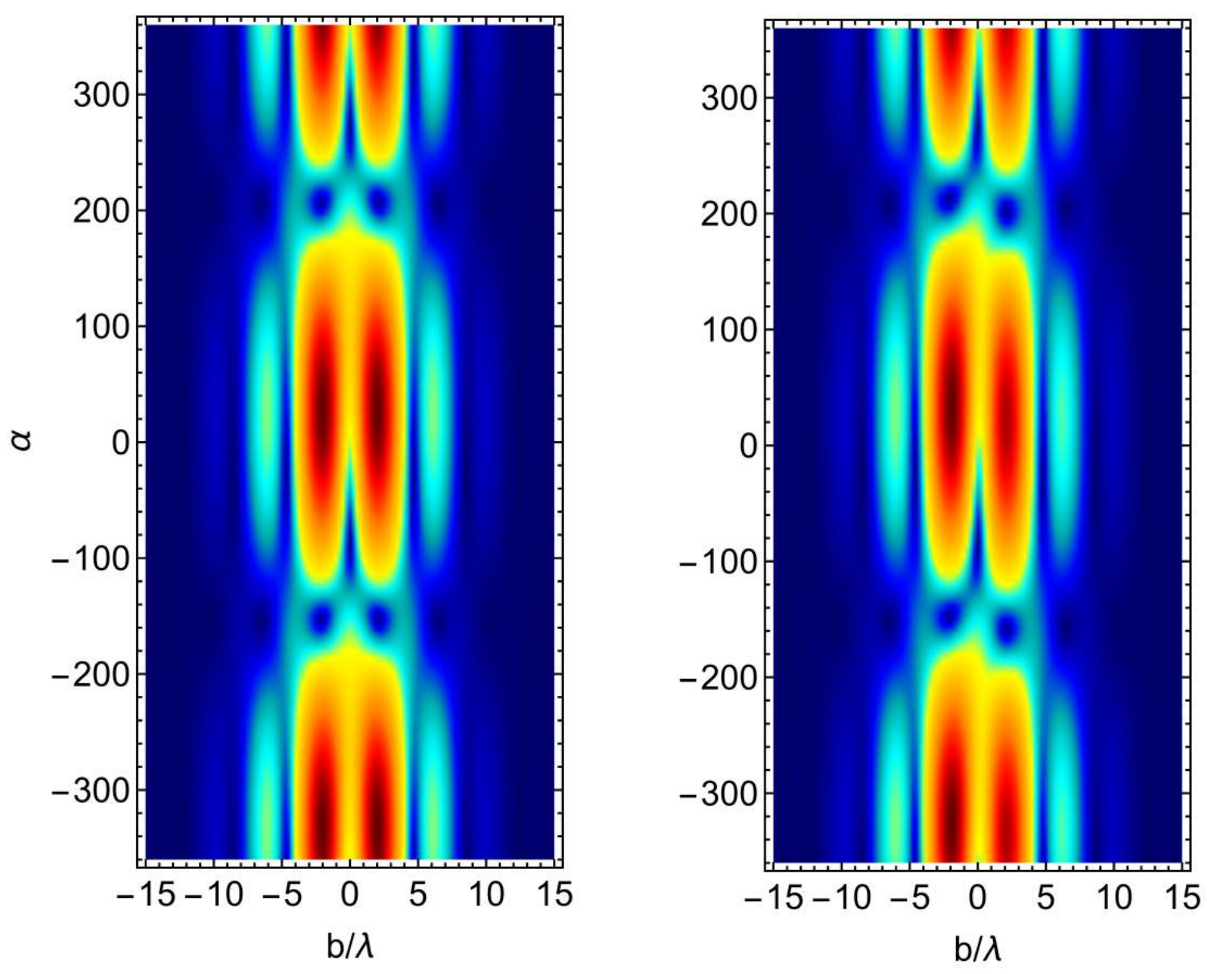}
\caption{(Colors online) Contour-plots of the normalized transition strength as a function of polarization and impact parameter for $^2S_{1/2} \rightarrow \;^2F_{7/2}$ in a single $^{172}\text{Yb}^+$. In both cases $\ell_{\gamma} = 1$, $\delta = 0$; but $\phi_b = 0$ ($-0.3$Rad) for left (right) plot.}
\label{08/01/2018_1}
\vspace{-4mm}
\end{figure}

For the case of linearly-polarized incoming photons, this can be rewritten as:
\begin{flalign}
M_{m_f m_i H/V}^{(\text{BG})} &(b; \theta_z) = \cos (\pi/4) M_{m_f, m_i \;-1}^{(\text{BG})} (b;\theta_z) - \sin (\pi/4) e^{-i2 \delta} M_{m_f, m_i \;1}^{(\text{BG})} (b;\theta_z) 
\label{06/06/2018_4}
\end{flalign}
Then we substitute the eqn. \eqref{01/10/18_1}, rotated by angle $\theta_z$, as in \eqref{01/24/2017/2}. The non-zero phase difference, coming from exponential factor $\exp [i(m_{\gamma} + m_i - m_f)\phi_b]$, introduces a new parameter $\phi_b$. In our simulations we are tuning this polar angle $\phi_b$, responsible for the azimuthal orientation of the ion with respect to the plane formed by the beam propagation axis and constant magnetic field. 

The calculation of the interaction strength involves two rotations performed sequentially: 1) the rotation of the plane wave photon by the pitch angle $\theta_k$ and 2) the rotation of the twisted beam over the angle $\theta_z$. While the $\phi_b$-dependent phase drops out on the level of Rabi frequencies $\Omega_R \propto |M_{m_f m_i \Lambda}|$ for definite helicity states, the same phase factor $\exp(i m_{\gamma} \phi_b)$ becomes observable for H and V-states. To control the polarization of the beam we used a set of wave-plates~\cite{Schmiegelow2016}. By rotating a half-wave-plate, the polarization state is varied along a meridian in the Poincar\'e sphere.

In Figure \ref{fig:plotsHandV}~(b), (e), (h) we show a comparison of the theoretical prediction with the experimental results for beams with different orbital angular momentum $\ell_{\gamma}=0,1,2$ with either vertical or horizontal polarization. Some features of the data can be understood by analytic expansions of~\eqref{29/09/17_2} for the various cases.   For small impact parameter $b \rightarrow 0$, the Bessel function will collapse into delta-function and the projection of the photon's total angular momentum $m_\gamma$ will be transferred into the internal degrees of freedom of the target atom, $\Delta m = m_f - m_i = m_{\gamma}$ \cite{Afanasev2013kaa, afanasev2016high, afanasev2018experimental}. In such a case the interaction strengths only depend on the pitch angle $\theta_k$ and the alignment angle $\theta_z$,
\begin{equation}
\begin{aligned}
&M^{\text{(BG)} (\ell_{\gamma} = 0) }_{3/2\;1/2\; H} \propto i  (5 \theta_k^2-4)\cos (2\theta_z )\;\;\;\;\;\;\;\;\;\;\;\;\;\;M^{\text{(BG)} (\ell_{\gamma} = 0)}_{3/2\;1/2\; V} \propto i  (5 \theta_k^2-4)\cos (\theta_z )\\
&M^{\text{(BG)} (\ell_{\gamma} = 1)}_{3/2\;1/2\; H} \propto  \; 2\theta_k (1+4 \cos \theta_z) \sin \theta_z\;\;\;\;\;\;\;M^{\text{(BG)} (\ell_{\gamma} = 1)}_{3/2\;1/2\; V} \propto  \; 2\theta_k (2 \cos \theta_z-1) \sin \theta_z\\
&M^{\text{(BG)} (\ell_{\gamma} = 2)}_{3/2\;1/2\; H} \propto i \; \frac{3}{\sqrt{2}}\theta_k^2 (\cos \theta_z + \cos 2\theta_z)\;\;\;\;\;M^{\text{(BG)} (\ell_{\gamma} = 2)}_{3/2\;1/2\; V} = -M^{\text{(tw)}}_{m_f\; m_i\; H}(\ell_{\gamma} = 2)
\label{eq:01/25/2018/4}
\end{aligned}
\end{equation}
From the equations above one can see that the horizontal polarization is completely suppressed in the vortex center when $\theta_z = \pi/4$ and $\ell_{\gamma} = 0$. In 45-degrees alignment this is a signature of a transitions with dominant $E2$. This is due to their sensitivity to the field gradients \cite{Schmiegelow2016} and generation of geometry-dependent terms due to rotation of the quantization axis \eqref{01/24/2017/2}. The dependence on alignment angle was previosly discussed in \cite{james1998quantum, roos2000controlling, schmiegelow2012light}.

When performing the expansion for different OAM, expressions in \eqref{eq:01/25/2018/4}, there is the crucial observation to make: the plane-wave transition amplitudes of different helicity do not contribute on equal basis as the OAM goes up. While for $\ell_{\gamma}=0, 1$  LC ($\Lambda = 1$) and RC ($\Lambda = -1$) contribute symmetrically, for $\ell_{\gamma} = 2$ only RC components contribute into the absorption matrix element at $b \rightarrow 0$. This leads to the the effect of local circular dichroism in isotropic targets.  This phenomenon is position dependent and gets stronger for beams with higher vorticity, until all the contributions become extinct at the vertex center. This result confirms and reinforces our earlier findings \cite{AfanasevJOPT17}.

In figure \ref{11/21/2017/1} we show the excitation profiles for a \textit{varying} polarization for a Gaussian beam and one with $\ell_{\gamma}=1$. The polarization state is scanned along a meridian in the Poincar\'e sphere parametrized by angle $\alpha$, as defined in eqn. \eqref{04/14/2018_1}. We see that the theoretical model reproduces all the main features, including the breaking of radial symmetry for polarizations which are not purely linear. The fit parameters are the beam waist $\text{w}_0$, the pitch angle $\theta_k$ and the azimuthal angle $\phi_b$.The fit was performed using the whole dataset. An overall normalization constant and pitch angle were picked for each profile independently, while the waist $\text{w}_0=9 \mu \text{m}/\lambda$ and azimuthal angle were constrained to be the same throughout. We obtain $\theta_z = 45^{\circ} \pm 5^{\circ}$ and $\delta =0\pm 0.02$. The polarization patterns are strongly dependent on angle $\phi_b$. In particular, when $\phi_b \rightarrow -\phi_b$ one would get a picture, mirror-symmetric to Figure \ref{11/21/2017/1}.

\begin{figure*}[!ht]
\begin{center}
\includegraphics[scale=0.29]{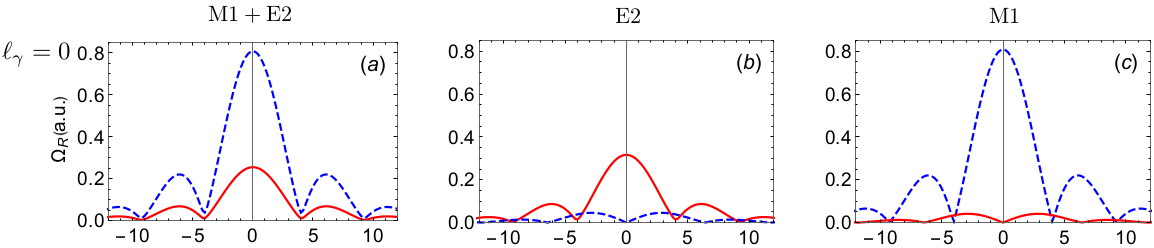}\\
\includegraphics[scale=0.29]{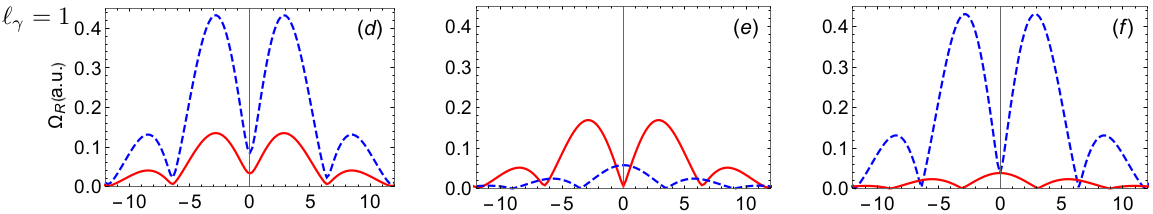}\\
\includegraphics[scale=0.29]{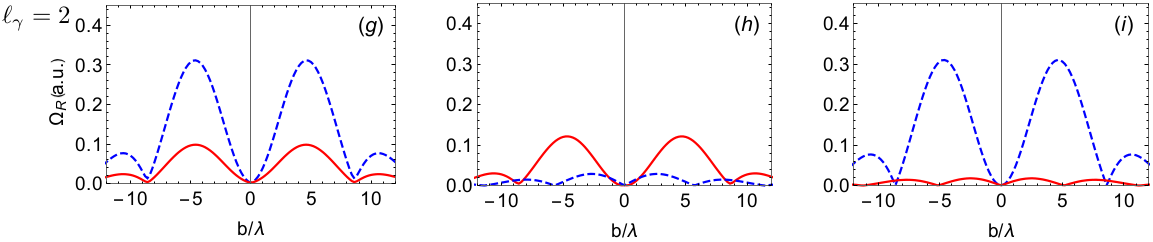}
\vspace{-0.2cm}
\caption{(Colors online) Transition strength $\Omega_r$ for different atomic multipolarities and beam types as a function of impact parameter $b$. The left column shows the expected response for a transition with similar contribution from the $M1$ and $E2$ multipoles such as for the HCI $^{20}\text{Ne}^{5+}$ on the transition  $^2P_{1/2} \rightarrow \;^2D_{3/2}$~\cite{rynkun2012energies}. Columns $E2$ and $M1$ show the individual contribution from each of the multipoles. 
In all the cases $\Delta m = 1$, $\phi_{b} = 0$ and the alignment angle is chosen to be $\theta_z = \pi/4$.
Each row shows the expected result for increasing orbital angular momentum $\ell_{\gamma} = 0,1,2$. 
Red-solid (blue-dashed) indicates results for light polarized vertically (horizontally) beams.}
\label{fig:plotsRandL}
\end{center}
\vspace{-0.8cm}
\end{figure*}

Finally in Figure~\ref{fig:plotsRandL}~(b), (e), (h) we show excitation profiles for alignment angle $\theta_z=0$ and left-circular polarization $\Lambda=1$.  We plot the results for different OAM $\ell_{\gamma}=0,1,2$ and different change in magnetic number $\Delta m=1,2,3$. At this angle the conservation of the projection of angular momentum for $b \rightarrow 0$ is observed. This is why, at the center of the beam the only non vanishing transitions are those in which the photon orbital angular momentum matches the change in magnetic number $\Delta m = m_{\gamma}$. However, it is interesting to note, that the $\Delta m =3$ transition is null even if the conservation conditions are met. This is due to an extra selection rule in which $|\Delta m|\leq|\Delta l|=2$ for $E2$ transitions.

\section{Predictions for $M1$, $E3$ and mixed multipolarity}

In this Section we present theoretical predictions of structured light field excitation for atomic ions featuring electric and magnetic multipole transitions of interest.

\subsection{$^{40}\text{Ar}^{13+}$ $M1$ - driven transition}

In the highly charged ion $^{40}\text{Ar}^{13+}$ the transition $P_{1/2} \rightarrow P_{3/2}$ is an example where the $M1$ amplitude dominates while the $E2$ is suppressed by six orders of magnitude \cite{fischer1983multiconfiguration}. The relevant plane-wave amplitudes from equation \eqref{10/5/17_1} are then
\vspace{-1mm}
\begin{equation}
\begin{gathered}
M_{(3\Lambda/2)\;(\Lambda/2)\;\Lambda}^{\text{(PW)}} = \Lambda \sqrt{3\pi} \; M1 \\ M_{(\Lambda/2)\;(-\Lambda/2)\;\Lambda}^{\text{(BB)}} = - \Lambda \sqrt{\pi} \; M1
\end{gathered}
\end{equation}
Following the procedure from the previous sections and under the condition of small $b$ one gets the transition strengths:
\vspace{-1mm}
\begin{equation}
\begin{aligned}
&M^{\text{(BG)} ( \ell_{\gamma} = 0 )}_{3/2\;1/2\;H} \propto -i(1- \frac{\theta_k^2}{4}) \;\;\;\;\;\;\;\;\;\;
M^{\text{(BG)}( \ell_{\gamma} = 0 )}_{3/2\;1/2\;V} \propto i(1 - \frac{\theta_k^2}{4}) \cos (\theta_z) \\
&M^{\text{(BG)} ( \ell_{\gamma} = 1 )}_{3/2\;1/2\;H/V} \propto   \theta_k\sin \theta_z \;\;\;\;\;\;\;\;\;\;\;\;
M^{\text{(BG)} ( \ell_{\gamma} = 2 )}_{3/2\;1/2\;H/V} \propto  \theta_k^2\cos^2 \frac{\theta_z}{2}
\label{eq:06/18/2018_1}
\end{aligned}
\end{equation}

For the case of the $\ell_{\gamma}=0$ and a setting of $\theta_z=0$ then both H and V interactions strengths are equal, this is identical to the behavior for $E1$ and $E2$ transitions. However, when varying $\theta_z$ to $\pi/2$ we see a distinct behavior: the value of the horizontally polarized beam does not change while the vertical goes to zero. The behavior for $M3$ at $\pi/2$, see figure~\ref{fig:plotsHandV}~(a) is then similar to that of $E2$ at $\pi/4$, but with interchanged roles of the horizontal and vertical polarizations (see figure~\ref{fig:plotsHandV}). This can be understood by the fact the E and B vectors are orthogonal in a transverse electromagnetic wave. 
For higher orbital angular momentum another interesting feature is observed: the transition strengths are the same for H and V polarized beams independent of the alignment angle. This effect is a direct result of the circular dichroism, similar to Sec. \ref{sec.3}. This is exemplified in figure~\ref{fig:plotsHandV}~(d) where we see that for $b=0$ both transition strengths coincide. 
\subsection{$\text{Yb}^+$ isotopes and E3-driven transition}

In this subsection, we consider the electric octupole (E3) transition in ytterbium known for ultra precise optical frequency standards~\cite{huntemann2016single}. Ytterbium has several naturally abundant isotopes. This makes this element an interesting species to study the role of nuclear spin and OAM for the features of the excitation strength. Here, we will compare results for ions with a spin-less nucleus as in isotope 172 and with $\pmb{I}=1/2$ nuclear spin of isotope 171. The transition from the ground state $^2S_{1/2} \rightarrow ^2F_{7/2}$ is driven with light near 467~nm. In the case of $^{172}\text{Yb}^+$, where the nuclear spin is zero,  we can use the formalism that has been exploited in the case of $^{40}\text{Ca}^+$, and $^{40}\text{Ar}^{13+}$ transitions and will be used also for $^{20}\text{Ne}^{5+}$ in the section IV. C. The coupling scheme to use for $^{171}\text{Yb}^+$ is the one with the coefficients \eqref{06/06/2018_1} with the proper quantum number for angular momentum is $\pmb{F} = \pmb{J}+\pmb{I}$, not accounting for electric hyperfine interaction, \emph{e.g.} \cite{dzuba2016hyperfine}.

As we see in figure~\ref{fig:plotsHandV}, the behavior at the alignment angle $\theta_z=\pi/4$ is conceptually similar to the case of $E2$. The position of the minima and the relative amplitudes of each transition vary slightly. And there is also a dependence of the relative interaction strengths on the spin content of the nucleus. One can see that the ion's response does not change drastically due to the presence of the hyperfine splitting. The correction due to the presence of the nuclear spin is on the order of 10$\%$.

However, a striking and distinctive feature occurs for E3 transitions in the collinear case: $\theta_z=0$. As we show in figure~\ref{fig:plotsRandL1}, now we can drive $\Delta m=3$ transitions. These are particularly strong at the center of a the beam with $m_\gamma=3$ as expected from the full selection rules and shown in figure~\ref{fig:plotsRandL1}~(i).

The matrix elements at vortex center, as calculated for the previous transitions are:

\begin{equation}
\begin{aligned}
&M^{\text{(BG)($\ell_{\gamma} = 0$)}}_{1\;0\;H}  \propto i (4 - 11\theta_k^2) (\cos \theta_z + 15 \cos 3 \theta_z)\\
&M^{\text{(BG)($\ell_{\gamma} = 0$)}}_{1\;0\;V} \propto - i (4 - 11\theta_k^2) (3+ 5\cos 2\theta_z)\\
&M^{\text{(BG)($\ell_{\gamma} = 1$)}}_{1\;0\;H} \propto  - 4 \theta_k \;(23 + 20 \cos \theta_z + 45 \cos 2 \theta_z) \sin \theta_z\\
&M^{\text{(BG)($\ell_{\gamma} = 1)$}}_{1\;0\;V} \propto - 4 \theta_k\;(13 - 20 \cos \theta_z + 15 \cos 2 \theta_z) \sin \theta_z\\
&M^{\text{(BG)($\ell_{\gamma} = 2)$}}_{1\;0\;H} \propto i\frac{3}{2}\theta_k^2\;(22 + 7 \cos \theta_z + 10 \cos 2 \theta_z + 25 \cos 3 \theta_z)\\
&M^{\text{(BG)($\ell_{\gamma} = 2)$}}_{1\;0\;V} \propto i6 \theta_k^2\; (21 - 40 \cos \theta_z + 35 \cos 2 \theta_z) \cos^2 \frac{\theta_z}{2}
\end{aligned}
\end{equation}

Hence, the horizontal polarization is completely suppressed when impact parameter $b \rightarrow 0$ for $\theta_z = \pi/2$. Dichroism effects can be seen for $\ell_{\gamma}>2$ and the full amplitude extinction takes place for $\ell_{\gamma} \geq 5$ in the central region of the beam for both $^{171}\text{Yb}^+$ and $^{172}\text{Yb}^+$ isotopes.

Finally in Figure~\ref{08/01/2018_1}, we plot the profiles for varying polarizations at alignment angle $\theta_z=\pi/4$. It is remarkable how as small variation in polarization ($\Delta \phi_b = \pm 17^{\circ}$) results in strong visual effects of mirror symmetry breaking.

\subsection{$^{20}\text{Ne}^{5+}$ and $E2+M1$ separation}

In this section we consider $^2P_{1/2} \rightarrow \;^2D_{3/2}$ transitions in B-like $^{20}\text{Ne}^{5+}$ highly-charged-ion (HCI) at $\lambda = 142$~nm, where $M1/E2 \propto 1.1$  \cite{rynkun2012energies}.
As before, we use the multipole expansion of equation  \eqref{10/5/17_1} to get the regular contributions into the plane-wave absorption amplitude as
\begin{gather}
M^{\text{(pw)}}_{m_fm_i \Lambda} \propto \sqrt{5} C_{1/2\;m_i;\;2\;\Lambda}^{3/2\;m_f} E2 + \Lambda \sqrt{3}C_{1/2\;m_i;\;1\;\Lambda}^{3/2\;m_f} M1	
\label{05/06/2018}
\end{gather}
With the use of equations~\eqref{29/09/17_2},  \eqref{01/24/2017/2} and \eqref{10/5/17_1} one can work out the explicit form of expressions for the rotated transition amplitudes and predict the Rabi frequencies for this case. In the general case, the norm of the transition amplitude is not always free of interference terms. However, in this case, no such terms arise. For this reason the behavior for this transition is simply a linear combination of the ones for $M1$ and $E2$ as described above and shown in figure~\ref{fig:plotsHandV} and~\ref{fig:plotsRandL}.

For this HCI, we study the transition amplitudes for a horizontal and vertical polarizations, as shown in figure~\ref{fig:plotsRandL}. 
The resulting amplitude for a transition with a mixture of $M1$ and $E2$ is shown in the first column. The individual multipolar contributions are shown in second - $E2$ and third - $M1$ correspondingly. One can clearly see that the quadrupole contribution (second column) looks identical to the $E2$ transition discussed above for calcium ions (figure~\ref{11/21/2017/1}). As well as the magnetic dipole contribution (third column) is identical to $M1$ transition discussed above for argon HCIs.

Let's examine this behavior mathematically in proximity to the beam vortex in this case. For $\ell_{\gamma} = 0$ one gets:
\begin{equation}
\begin{aligned}
&M^{\text{(BG)}}_{3/2\;1/2\;H} \propto i(\sqrt{3} M1 - E2 (1-2\cos \theta_k) \cos 2\theta_z)\\
&M^{\text{(BG)}}_{3/2\;1/2\;V} \propto i(\sqrt{3} M1 - E2 (1-2\cos \theta_k) \cos \theta_z) \label{01/25/2018/2}
\end{aligned}
\end{equation}


For $M1$ $\rightarrow0$ we get the same approximation as in the purely $E2$-driven case \eqref{eq:01/25/2018/4}. In the same way, $E2$ $\rightarrow 0$ gives eqn.\eqref{eq:06/18/2018_1}. The same can be proved for higher values of OAM $\ell_{\gamma}$. This means that for the case with mixed multipolarity the horizontal component cannot be completely suppressed at any $\alpha$, unless magnetic contribution is negligible or the pitch angle $\theta_k$ is very small. Second, however, would mean taking the limit of OAM going to zero - the plane-wave limit. Hence, presence of strong $M1$ contribution sensibly changes the photo-absorption behavior at the beam center.

Both $^{20}\text{Ne}^{5+}$ and $^{40}\text{Ar}^{13+}$ have first non-zero contribution coming from a magnetic dipole. The observed in figure~\ref{fig:plotsRandL} (a,d,g) drastic difference in neon optical response in the area of the beam penumbra, compared to argon figure~\ref{fig:plotsHandV} (a,d,g), is due to the presence of the second strongest multipole - electric quadrupole. It can be inferred that contributions coming from higher order multipoles may result into visible effects in the central region, as in this case, turning minima into maxima in position dependence of Rabi frequency. This would be especially useful for fundamental studies in spectroscopy of highly charged ions. 

\section{Conclusions and Outlook}

In this paper, we have studied the modifications of selection rules for electric and magnetic transitions, particularly $E2$, $E3$, and $M1$, when well localized single ions interact with focused tailored electromagnetic fields with vortices. We have developed a complete theoretical framework, extending previous results, \textit{e.g.}~\cite{afanasev2018experimental}, where we can determine the full dependence of the photoabsorption transition amplitudes on orbital angular momentum, on polarization of the beam, on impact parameter, on nuclear spin when relevant, and on directions of the quantization axis of the atoms.  This whole framework for single ion -- vortex interactions is tested and verified using experimental data for 
$^{40}$Ca$^+$~\cite{Schmiegelow2016}.

We also  encourage development of future experiments involving twisted photon photoaborption on various ions, including Yb (used for clocks) and other HCI, by predicting rates for selected transitions and a variety of structured electromagnetic beams.    We highlight peculiar effects, which can form a rich ground of fundamental studies, and will \textit{e.g.}~allow separating the E and M character for given atomic transitions, and enhance the excitation rates in certain circumstances.

The calculations give transition amplitudes, which for fixed laser strength are equivalent to Rabi oscillation frequencies, as experimentally measured.  The experimental results for the $^{40}$Ca ion which confirmed the validity of the calculations were for the case with a 45$^{\circ}$ static magnetic field alignment and in the full polarization domain of the laser beam for OAM $\ell_{\gamma} = 0$ and $1$.   Together with the predictions in related configurations for Ar, Ne, and Yb ions, one may conclude that local, or small impact parameter, alteration of the ordinary spectroscopic selection rules enables selective enhancement of high-order multipolar contributions in the ion response by manipulating the photon twist. Circular dichroism related to the twisted photon topology is theoretically seen on the level of factorized multipolar contributions.   The reported circular dichroism in atom-photon interactions may contribute into the study of chiral light-matter interactions \cite{lodahl2017chiral}.

The ions and transitions studied in this paper are excited by different multipolarities.  The results from horizontal and vertical linear polarization states become most distinct at particular static magnetic field alignment angles $\theta_z$, with the angle of most sensitivity depending on the photon-atom angular momentum transfer.  For the system with $M1$ transitions, or partial wave amplitude with $j=1$, choose $\theta_z = \pi/2$;  for $E2$ or $j=2$, choose $\theta_z = \pi/4$; and for E3 or $j=3$, choose again $\theta_z = \pi/2$.  For the Boron-like $^{20}\text{Ne}^{5+}$ HCI, the $M1$ and $E2$ partial waves are unusually similar, leading to effects shown in this paper;  see also~\cite{afanasev2018E2M1}.   This makes this HCI a promising candidate for future experimental research of topological effects in light-matter interactions. Progress in this direction will also allow further investigation of the quantum nature of spin-orbit coupling in photon laser beams.

Also included are a number of results for $\theta_z=0$, with transition amplitudes for differing OAM $\ell_\gamma$ with positive circular polarization $\Lambda$, and with differing changes in the atomic angular momentum projections, $\Delta m$.  One sees again~\cite{Schmiegelow2016,afanasev2018experimental} the non-zero response at zero impact parameter only for $\Delta m = m_\gamma = \ell_\gamma+\Lambda$, and of note is the good agreement with experimental data available in a range of impact parameters for one of the $^{40}$Ca$^+$ cases.  

Photon OAM coupling to the internal atomic degrees of freedom can be used, as seen here, to separate the electric and magnetic character of given transitions, to locally enhance next-to-leading order transition rates and to suppress parasitic transitions. In the realm of quantum computation, incorporation of OAM states of light can provide ion-photon interaction on the level of energy, momentum, polarization and phase.

The theory outlined in this paper can also be successfully applied to such systems as bulk semiconductors and artificial atoms in the ways, similar to \cite{quinteiro2009theory, solyanik2017interband}. When developed, it may become useful in such fields as metrology, classical and quantum communication, quantum computing, high capacity data transfer, cybersecurity to name a few. The highlighted sensitivity to the target position in the beam, polarization and phase can be used in beam diagnostics.

\section{Funding Information}

National Science Foundation (NSF) (1516509, 1812326); Gus Weiss Endowment of George Washington University.

\section{Acknowledgement}

C.E.C thanks the Johannes Gutenberg-University Mainz for hospitality while this work was underway.   The authors would like to thank Valery Serbo and Dmitry Budker for useful discussions. Special thanks to Nicholas Gorgone for his help in working on this manuscript.

\section{Appendix}

One might benefit from linking two alternative sets of angles $\{ \pm \psi_k, \pm \theta_k, \pm \phi_k \}$, equivalent to active and passive rotation 3D rotation. The conventional formalism \cite{akhiezer1959quantum, afanasev2018E2M1} for photo-absorption of the plane wave of arbitrary incidence angle $\Omega$ by the atomic system may be written as

\begin{equation}
\begin{aligned}
M_{m_f m_i \Lambda}^{(\text{pw})} (\Omega) =& -\sqrt{4\pi} \sum_{j=1}^{\infty} \sum_{m=-j}^{j} i^{j+\mu} \sqrt{\frac{2j+1}{2j_f + 1}} \Lambda^{\mu + 1} \times && \\ &\times D_{\Lambda m}^{j\;*} (\psi_k, \theta_k, 0) (-1)^{j-j_f+j_i} C_{j_i\;m_i\;j\; m}^{^{j_f\;m_f}}M_{j \mu}&&
\label{24/07/2018_2}
\end{aligned}
\end{equation}
where the rotation matrix is related to ours as $D_{mm'}^{j} (\psi_k, \theta_k, \phi_k)= e^{-i m \psi_k} d_{mm'}^{j} (\theta_k) e^{-i m' \phi_k}$. This matrix represents active Euler rotation in Hilbert space. The part of the tensor, responsible for the spatial configuration of the system

\begin{equation}
\Lambda D_{\Lambda m}^{j\;*} (\psi_k, \theta_k, 0) C_{j_i\;m_i\;j\; m}^{^{j_f\;m_f}}
\label{24/07/2018_1}
\end{equation}
reproduces up to a sign the one in semi-classical formalism \cite{james1998quantum, roos2000controlling}. It can be checked in direct calculation and the results are presented in the top part of Table \ref{tab1}.

If, instead of rotating the photon state, we rotate an electron state, similar to \eqref{29/09/17_2}:

\begin{equation}
\begin{aligned}
M_{m_f m_i \Lambda}^{(\text{pw})} (\Omega) =& -\sqrt{4\pi} \sum_{j=1}^{\infty} \sum_{m'_f, m'_i} i^{j+\mu} \sqrt{\frac{2j+1}{2j_f + 1}} \Lambda^{\mu + 1} \times && \\ &\times D_{m_f m'_f}^{j_f} (0, -\theta_k, -\psi_k) D_{m_i m'_i}^{j_i\;*} (0, -\theta_k, -\psi_k) (-1)^{j-j_f+j_i} C_{j_i\;m'_i\;j\; \Lambda}^{^{j_f\;m'_f}}M_{j \mu}&&
\end{aligned}
\end{equation}

\begin{center}
\begin{table*}[ht!]
  \centering
  \caption{The geometry-dependent terms in the plane-wave photo-absorption matrices \eqref{24/07/2018_2} for two sets of Euler angles: top - $\{ \psi_k, \theta_k, 0 \}$, and bottom - $\{ 0, -\theta_k, -\phi_k \}$}
  \scalebox{0.8}{
  \begin{tabular}{|P{1.8cm}|P{2.2cm}|P{5.5cm}| P{5.5cm}|}
    \hline
    \diagbox{$H/V$}{$\Delta m$} & 0 & $\pm1$ & $\pm2$ \\ \hline
    H & $\sin (2\theta_k) \cos \psi_k$ & $ \pm \cos 2 \theta_k \cos \psi_k - i \cos \theta_k \sin \psi_k $ & $ - \frac{1}{2}\sin(2\theta_k) \cos \psi_k \pm i \sin \theta_k \sin \psi_k$ \\ \hline
    V & $i\sin (2\theta_k) \sin \psi_k$ & $-\cos \theta_k \cos \psi_k \pm i \cos(2 \theta_k)\sin \psi_k$ & $\pm \sin \theta_k \cos \psi_k - i \frac{1}{2} \sin(2\theta_k) \sin \psi_k$ \\ \hline \hline
    H & $-\sin (2\theta_k)$ & $ \pm \cos (2 \theta_k) \cos \phi_k + i \cos (2\theta_k) \sin \phi_k$ & $\sin (2\theta_k) \cos (2\phi_k) \pm i \sin (2 \theta_k) \sin (2\phi_k)$ \\ \hline
    V & 0 &$-\cos \theta_k (\cos \phi_k \pm i \sin \phi_k)$ & $-\sin \theta_k ( \pm \cos (2\phi_k) + i \sin (2\phi_k))$ \\ \hline
  \end{tabular}
  }
  \label{tab1}
\end{table*}
\end{center}
\FloatBarrier

This is the passive rotation $\{0, -\theta_k, -\psi_k\}$, equivalent to $\{\psi_k, \theta_k, 0\}$. The configuration-dependent part of this equation 

\begin{equation}
\begin{aligned}
\sum_{m'_f, m'_i} & \Lambda C_{j_i\;m'_i\;j\; \Lambda}^{^{j_f\;m'_f}} D_{m_f m'_f}^{j_f} (0, -\theta_k, -\psi_k) D_{m_i m'_i}^{j_i\;*} (0, -\theta_k, -\psi_k)&&
\end{aligned}
\end{equation}
is identical to \eqref{24/07/2018_1}, though containing an extra Wigner rotation matrix. In our formalism, Sec. \ref{Sec.2}, we have used the rotation system, defined in terms of the Euler angles in \eqref{24/07/2018_2} as $\{ 0, -\theta_k, -\phi_k \}$. The resulting geometry-related terms are presented in the bottom half of the Table \ref{tab1}. If we proceed with this alternative description in terms of active rotation, we get the equation, analogous to \eqref{29/09/17_2},
\begin{equation}
\begin{aligned}
M_{m_f m_i \Lambda}^{(\text{BB})} (b; \theta_z &=0) = A \sum_{m} i^{m-2m_{\gamma}} e^{i(m_{\gamma} - m) \phi_k} J_{m_{\gamma} - m} (\kappa b) d_{m\Lambda}^{j}(\theta_k) M_{m_fm_i \Lambda}^{\text{(pw)}}(0)
\end{aligned}
\end{equation}
where the plane-wave amplitude is defined as \eqref{10/5/17_1} with the coupling coefficient being

\begin{equation}
C_{jj_fj_i}^{\Lambda} = (-1)^{j-j_f + j_i} C_{j_i, m_i, j, m}^{j_f, m_f}
\label{10/03/2018_1}
\end{equation}
As one can see, this way photon TAM projection $m_{\gamma}$ explicitly dictates multipolar balance in optical response of the ion in the beam penumbra, as discussed in \cite{schmiegelow2012light}.



\bibliography{Master}

\begin{thebibliography}{10}
\newcommand{\enquote}[1]{``#1''}

\bibitem{nye1974dislocations}
J.~F. Nye and M.~V. Berry, \enquote{Dislocations in wave trains,}
  {\protect\JournalTitle{Proc. R. Soc. Lond. A}} \textbf{336}, 165--190 (1974).

\bibitem{wesfreid1984cellular}
J.~E. Wesfreid and S.~Zaleski, \enquote{Cellular structures in instabilities,}
  in \emph{Cellular Structures in Instabilities,}  vol. 210 (1984).

\bibitem{1464-4258-11-9-090201}
M.~R. Dennis, Y.~S. Kivshar, M.~S. Soskin, and G.~A.~S. Jr, \enquote{Singular
  {O}ptics: more ado about nothing,} {\protect\JournalTitle{Journal of Optics
  A: Pure and Applied Optics}} \textbf{11}, 090201 (2009).

\bibitem{yao2011orbital}
A.~M. Yao and M.~J. Padgett, \enquote{Orbital angular momentum: origins,
  behavior and applications,} {\protect\JournalTitle{Adv. Opt. Photon.}}
  \textbf{3}, 161--204 (2011).

\bibitem{Allen1992zz}
L.~Allen, M.~Beijersbergen, R.~Spreeuw, and J.~Woerdman, \enquote{{Orbital
  angular momentum of light and the transformation of {L}aguerre-{G}aussian
  laser modes},} {\protect\JournalTitle{Phys.Rev.}} \textbf{A45}, 8185--8189
  (1992).

\bibitem{van1994spin}
S.~Van~Enk and G.~Nienhuis, \enquote{Spin and orbital angular momentum of
  photons,} {\protect\JournalTitle{EPL}} \textbf{25}, 497 (1994).

\bibitem{sztul2006double}
H.~Sztul and R.~Alfano, \enquote{Double-slit interference with
  {L}aguerre-{G}aussian beams,} {\protect\JournalTitle{Opt. Lett.}}
  \textbf{31}, 999--1001 (2006).

\bibitem{bliokh2015spin}
K.~Y. Bliokh, F.~Rodr{\'\i}guez-Fortu{\~n}o, F.~Nori, and A.~V. Zayats,
  \enquote{Spin--orbit interactions of light,} {\protect\JournalTitle{Nat.
  Photonics}} \textbf{9}, 796 (2015).

\bibitem{milione2011higher}
G.~Milione, H.~Sztul, D.~Nolan, and R.~Alfano, \enquote{Higher-order
  {P}oincar{\'e} sphere, {S}tokes parameters, and the angular momentum of
  light,} {\protect\JournalTitle{Phys. Rev. Lett.}} \textbf{107}, 053601
  (2011).

\bibitem{Surzhikov15}
A.~Surzhykov, D.~Seipt, V.~G. Serbo, and S.~Fritzsche, \enquote{Interaction of
  twisted light with many-electron atoms and ions,}
  {\protect\JournalTitle{Phys. Rev. A}} \textbf{91}, 013403 (2015).

\bibitem{Schmiegelow2016}
C.~T. Schmiegelow, J.~Schulz, H.~Kaufmann, T.~Ruster, U.~G. Poschinger, and
  F.~Schmidt-Kaler, \enquote{Transfer of optical orbital angular momentum to a
  bound electron,} {\protect\JournalTitle{Nat. Comm.}} \textbf{7}, 12998
  (2016).

\bibitem{quinteiro2017twisted}
G.~F. Quinteiro, F.~Schmidt-Kaler, and C.~T. Schmiegelow,
  \enquote{Twisted-{L}ight--{I}on {I}nteraction: {T}he {R}ole of {L}ongitudinal
  {F}ields,} {\protect\JournalTitle{Phys. Rev. Lett.}} \textbf{119}, 253203
  (2017).

\bibitem{padgett2017orbital}
M.~J. Padgett, \enquote{Orbital angular momentum 25 years on,}
  {\protect\JournalTitle{Opt. Express}} \textbf{25}, 11265--11274 (2017).

\bibitem{brunet2016optical}
C.~Brunet and L.~A. Rusch, \enquote{Optical fibers for the transmission of
  orbital angular momentum modes,} {\protect\JournalTitle{Opt. Fiber Technol.}}
  \textbf{31}, 172--177 (2016).

\bibitem{krenn2016twisted}
M.~Krenn, J.~Handsteiner, M.~Fink, R.~Fickler, R.~Ursin, M.~Malik, and
  A.~Zeilinger, \enquote{Twisted light transmission over 143 km,}
  {\protect\JournalTitle{Proc. Natl. Acad. Sci. USA}} \textbf{113},
  13648--13653 (2016).

\bibitem{mair2001entanglement}
A.~Mair, A.~Vaziri, G.~Weihs, and A.~Zeilinger, \enquote{Entanglement of the
  orbital angular momentum states of photons,} {\protect\JournalTitle{Nature}}
  \textbf{412}, 313 (2001).

\bibitem{fickler2012quantum}
R.~Fickler, R.~Lapkiewicz, W.~N. Plick, M.~Krenn, C.~Schaeff, S.~Ramelow, and
  A.~Zeilinger, \enquote{Quantum entanglement of high angular momenta,}
  {\protect\JournalTitle{Science}} \textbf{338}, 640--643 (2012).

\bibitem{krenn2017orbital}
M.~Krenn, M.~Malik, M.~Erhard, and A.~Zeilinger, \enquote{Orbital angular
  momentum of photons and the entanglement of {L}aguerre--{G}aussian modes,}
  {\protect\JournalTitle{Phil. Trans. R. Soc. A}} \textbf{375}, 20150442
  (2017).

\bibitem{monticone2017metamaterial}
F.~Monticone and A.~Al{\`u}, \enquote{Metamaterial, plasmonic and nanophotonic
  devices,} {\protect\JournalTitle{Rep. Prog. Phys.}} \textbf{80}, 036401
  (2017).

\bibitem{swartzlander2001peering}
G.~A. Swartzlander, \enquote{Peering into darkness with a vortex spatial
  filter,} {\protect\JournalTitle{Opt. Lett.}} \textbf{26}, 497--499 (2001).

\bibitem{maurer2011spatial}
C.~Maurer, A.~Jesacher, S.~Bernet, and M.~Ritsch-Marte, \enquote{What spatial
  light modulators can do for optical microscopy,} {\protect\JournalTitle{Laser
  Photonics Rev.}} \textbf{5}, 81--101 (2011).

\bibitem{trichili2016optical}
A.~Trichili, C.~Rosales-Guzm{\'a}n, A.~Dudley, B.~Ndagano, A.~B. Salem,
  M.~Zghal, and A.~Forbes, \enquote{Optical communication beyond orbital
  angular momentum,} {\protect\JournalTitle{Sci. Rep}} \textbf{6}, 27674
  (2016).

\bibitem{Yao11}
A.~Yao and M.~Padgett, \enquote{Orbital angular momentum: Origins, behavior and
  applications,} {\protect\JournalTitle{Adv. Opt. Photon.}} \textbf{3},
  161--204 (2011).

\bibitem{barnett2016optical}
S.~M. Barnett, L.~Allen, and M.~J. Padgett, \emph{Optical angular momentum}
  (CRC Press, 2016).

\bibitem{Padgett2015}
E.~Wisniewski-Barker and M.~Padgett, \enquote{Orbital angular momentum,}
  {\protect\JournalTitle{Photonics: Scientific Foundations, Technology and
  Applications}} \textbf{1}, 321--340 (2015).

\bibitem{veysi2016focused}
M.~Veysi, C.~Guclu, and F.~Capolino, \enquote{Focused azimuthally polarized
  vector beam and spatial magnetic resolution below the diffraction limit,}
  {\protect\JournalTitle{JOSA B}} \textbf{33}, 2265--2277 (2016).

\bibitem{quinteiro2017formulation}
G.~Quinteiro, D.~Reiter, and T.~Kuhn, \enquote{Formulation of the
  twisted-light--matter interaction at the phase singularity: Beams with strong
  magnetic fields,} {\protect\JournalTitle{Phys. Rev. A}} \textbf{95}, 012106
  (2017).

\bibitem{babiker2002orbital}
M.~Babiker, C.~Bennett, D.~Andrews, and L.~D. Romero, \enquote{Orbital angular
  momentum exchange in the interaction of twisted light with molecules,}
  {\protect\JournalTitle{Phys. Rev. Lett.}} \textbf{89}, 143601 (2002).

\bibitem{scholz2014absorption}
H.~Scholz-Marggraf, S.~Fritzsche, V.~Serbo, A.~Afanasev, and A.~Surzhykov,
  \enquote{Absorption of twisted light by hydrogenlike atoms,}
  {\protect\JournalTitle{Phys. Rev. A}} \textbf{90}, 013425 (2014).

\bibitem{quinteiro2010electronic}
G.~Quinteiro, A.~Lucero, and P.~Tamborenea, \enquote{Electronic transitions in
  quantum dots and rings induced by inhomogeneous off-centered light beams,}
  {\protect\JournalTitle{J. Phys.: Condens. Matter}} \textbf{22}, 505802
  (2010).

\bibitem{quinteiro2017magnetic}
G.~F. Quinteiro, D.~Reiter, and T.~Kuhn, \enquote{Magnetic-optical transitions
  induced by twisted light in quantum dots,} in \emph{J. Phys.: Conf. Ser.},
  vol. 906. No. 1. (IOP Publishing, 2017), p. 012014.

\bibitem{afanasev2018E2M1}
A.~Afanasev, C.~E. Carlson, and M.~Solyanik, \enquote{Atomic spectroscopy with
  twisted photons: Separation of {M1$\ensuremath{-}$E2} mixed multipoles,}
  {\protect\JournalTitle{Phys. Rev. A}} \textbf{97}, 023422 (2018).

\bibitem{van1994selection}
S.~Van~Enk, \enquote{Selection rules and centre-of-mass motion of ultracold
  atoms,} {\protect\JournalTitle{Quantum Opt.}} \textbf{6}, 445 (1994).

\bibitem{franke2017optical}
S.~Franke-Arnold, \enquote{Optical angular momentum and atoms,}
  {\protect\JournalTitle{Phil. Trans. R. Soc. A}} \textbf{375}, 20150435
  (2017).

\bibitem{peshkov2017photoexcitation}
A.~Peshkov, D.~Seipt, A.~Surzhykov, and S.~Fritzsche, \enquote{Photoexcitation
  of atoms by {L}aguerre-{G}aussian beams,} {\protect\JournalTitle{Phys. Rev.
  A}} \textbf{96}, 023407 (2017).

\bibitem{afanasev2018experimental}
A.~Afanasev, C.~E. Carlson, C.~Schmiegelow, J.~Schulz, F.~Schmidt-Kaler, and
  M.~Solyanik, \enquote{Experimental verification of position-dependent
  angular-momentum selection rules for absorption of twisted light by a bound
  electron,} {\protect\JournalTitle{New J. Phys.}} \textbf{20}, 023032 (2018).

\bibitem{afanasev2016high}
A.~Afanasev, C.~E. Carlson, and A.~Mukherjee, \enquote{High-multipole
  excitations of hydrogen-like atoms by twisted photons near a phase
  singularity,} {\protect\JournalTitle{J. Opt.}} \textbf{18}, 074013 (2016).

\bibitem{haffner2008quantum}
H.~H{\"a}ffner, C.~F. Roos, and R.~Blatt, \enquote{Quantum computing with
  trapped ions,} {\protect\JournalTitle{Phys. Rep}} \textbf{469}, 155--203
  (2008).

\bibitem{schmoger2015coulomb}
L.~Schm{\"o}ger, O.~Versolato, M.~Schwarz, M.~Kohnen, A.~Windberger, B.~Piest,
  S.~Feuchtenbeiner, J.~Pedregosa-Gutierrez, T.~Leopold, P.~Micke
  \emph{et~al.}, \enquote{Coulomb crystallization of highly charged ions,}
  {\protect\JournalTitle{Science}} \textbf{347}, 1233--1236 (2015).

\bibitem{lu2000commercial}
P.~Lu, H.~Nakano, T.~Nishikawa, and N.~Uesugi, \enquote{Commercial terawatt
  femtosecond laser-driven tabletop {X}-ray lasers in gases,} in \emph{Proc.
  SPIE,}  vol. 3886 (International Society for Optics and Photonics, 2000), pp.
  294--306.

\bibitem{chen2004experimental}
H.~Chen, P.~Beiersdorfer, L.~Heeter, D.~Liedahl, K.~Naranjo-Rivera,
  E.~Tr{\"a}bert, M.~Gu, and J.~Lepson, \enquote{Experimental and theoretical
  evaluation of density-sensitive {N VI}, {Ar XIV}, and {Fe XXII} line ratios,}
  {\protect\JournalTitle{ApJ}} \textbf{611}, 598 (2004).

\bibitem{olmschenk2007manipulation}
S.~Olmschenk, K.~Younge, D.~Moehring, D.~Matsukevich, P.~Maunz, and C.~Monroe,
  \enquote{Manipulation and detection of a trapped {Yb$^+$} hyperfine qubit,}
  {\protect\JournalTitle{Phys. Rev. A}} \textbf{76}, 052314 (2007).

\bibitem{welinski2016high}
S.~Welinski, A.~Ferrier, M.~Afzelius, and P.~Goldner, \enquote{High-resolution
  optical spectroscopy and magnetic properties of {Yb $^{3+}$} in
  {Y$_2$}{SiO$_5$},} {\protect\JournalTitle{Phys. Rev. B}} \textbf{94}, 155116
  (2016).

\bibitem{ortu2018simultaneous}
A.~Ortu, A.~Tiranov, S.~Welinski, F.~Fr{\"o}wis, N.~Gisin, A.~Ferrier,
  P.~Goldner, and M.~Afzelius, \enquote{Simultaneous coherence enhancement of
  optical and microwave transitions in solid-state electronic spins,}
  {\protect\JournalTitle{Nat. Mater.}} \textbf{17}, 671 (2018).

\bibitem{wcis?o2018First}
P.~Wcislo, P.~Ablewski, K.~Beloy, S.~Bilicki, M.~Bober, R.~Brown, R.~Fasano,
  R.~Ciurylo, and H.~Hachisu, \enquote{First observation with global network of
  optical atomic clocks aimed for a dark matter detection,}
  {\protect\JournalTitle{arXiv:1806.04762}}  (2018).

\bibitem{rynkun2012energies}
P.~Rynkun, P.~J{\"o}nsson, G.~Gaigalas, and C.~F. Fischer, \enquote{Energies
  and {E1, M1, E2, M2} transition rates for states of the $2s^22p$, $2s^2p2$,
  and $2p^3$ configurations in boron-like ions between {N III} and {Zn XXVI},}
  {\protect\JournalTitle{At. Data Nucl. Data Tables}} \textbf{98}, 481--556
  (2012).

\bibitem{varshalovich1988quantum}
D.~Varshalovich, A.~Moskalev, and V.~Khersonskii, \emph{Quantum theory of
  angular momentum} (World Scientific, 1988).

\bibitem{siegman1986university}
A.~E. Siegman, \emph{Lasers}, vol.~37 (University Science Books, Mill Valley,
  CA, 1986).

\bibitem{rodrigues2016excitation}
J.~Rodrigues, L.~Marcassa, and J.~Mendon{\c{c}}a, \enquote{Excitation of high
  orbital angular momentum {R}ydberg states with {L}aguerre--{G}auss beams,}
  {\protect\JournalTitle{J. Phys. B: Atomic, Molecular and Optical Physics}}
  \textbf{49}, 074007 (2016).

\bibitem{edmonds1957angular}
A.~Edmonds, \enquote{Angular momentum in quantum mechanics,}  (1957).

\bibitem{condon1951theory}
E.~U. Condon, E.~Condon, and G.~H. Shortley, \emph{The theory of atomic
  spectra} (Cambridge University Press, 1951).

\bibitem{Afanasev2013kaa}
A.~Afanasev, C.~E. Carlson, and A.~Mukherjee, \enquote{Off-axis excitation of
  hydrogenlike atoms by twisted photons,} {\protect\JournalTitle{Phys. Rev. A}}
  \textbf{88}, 033841 (2013).

\bibitem{james1998quantum}
D.~F. James, \enquote{Quantum dynamics of cold trapped ions with application to
  quantum computation,} {\protect\JournalTitle{App. Phys. B}} \textbf{66},
  181--190 (1998).

\bibitem{roos2000controlling}
C.~Roos, \enquote{Controlling the quantum state of trapped ions,}
  {\protect\JournalTitle{Graz Karl-Franzens-Univ, PhD thesis}}  (2000).

\bibitem{schmiegelow2012light}
C.~T. Schmiegelow and F.~Schmidt-Kaler, \enquote{Light with orbital angular
  momentum interacting with trapped ions,} {\protect\JournalTitle{Eur. Phys. J.
  D}} \textbf{66}, 157 (2012).

\bibitem{AfanasevJOPT17}
A.~Afanasev, C.~E. Carlson, and M.~Solyanik, \enquote{Circular dichroism of
  twisted photons in non-chiral atomic matter,} {\protect\JournalTitle{J.
  Opt.}} \textbf{19}, 105401 (2017).

\bibitem{fischer1983multiconfiguration}
C.~F. Fischer, \enquote{Multiconfiguration {H}artree-{F}ock {B}reit-{P}auli
  results for 2{P}$_{1/2}$-2{P}$_{3/2}$ transitions in the {B}oron sequence,}
  {\protect\JournalTitle{J. Phys. B: Atomic and Molecular Physics}}
  \textbf{16}, 157 (1983).

\bibitem{huntemann2016single}
N.~Huntemann, C.~Sanner, B.~Lipphardt, C.~Tamm, and E.~Peik,
  \enquote{Single-ion atomic clock with 3$\times10^{18}$ systematic
  uncertainty,} {\protect\JournalTitle{Phys. Rev. Lett.}} \textbf{116}, 063001
  (2016).

\bibitem{dzuba2016hyperfine}
V.~Dzuba and V.~Flambaum, \enquote{Hyperfine-induced electric dipole
  contributions to the electric octupole and magnetic quadrupole atomic clock
  transitions,} {\protect\JournalTitle{Phys. Rev. A}} \textbf{93}, 052517
  (2016).

\bibitem{lodahl2017chiral}
P.~Lodahl, S.~Mahmoodian, S.~Stobbe, A.~Rauschenbeutel, P.~Schneeweiss,
  J.~Volz, H.~Pichler, and P.~Zoller, \enquote{Chiral quantum optics,}
  {\protect\JournalTitle{Nature}} \textbf{541}, 473 (2017).

\bibitem{quinteiro2009theory}
G.~Quinteiro and P.~Tamborenea, \enquote{Theory of the optical absorption of
  light carrying orbital angular momentum by semiconductors,}
  {\protect\JournalTitle{EPL (Europhys. Lett.)}} \textbf{85}, 47001 (2009).

\bibitem{solyanik2017interband}
M.~Solyanik and A.~Afanasev, \enquote{Interband absorption of topologically
  structured photon beams by semiconducting quantum dots,} in \emph{Laser
  Science,}  (Optical Society of America, 2017), pp. JTu3A--79.

\bibitem{akhiezer1959quantum}
A.~Akhiezer and V.~Berestetskii, \enquote{Quantum electrodynamics,}
  {\protect\JournalTitle{Interscience, New York}}  (1959).

\end{thebibliography}

\end{document}